    \documentstyle[epsf]{article} \catcode`\@=11
    \def\section{\@startsection{section}{1}{\z@}%
    {-3.5ex plus -1ex minus -.5ex}{1.5ex plus.3ex}{\bf }}
    \def\subsection{\@startsection{subsection}{1}{\z@}%
    {-3.5ex plus-1ex minus-.5ex}{1.5ex plus.3ex}{\bf }} 
    
\begin{document}
    \hfill\parbox{4.77cm}{\Large\centering Annalen\\der
    Physik\\[-.2\baselineskip] {\small \underline{\copyright\ Johann
    Ambrosius Barth 1998}}}
    \vspace{.75cm}\newline{\Large\bf
      First Order Metal-Insulator Transition \\
      in Two-dimensional Disordered Systems}
    \vspace{.4cm}\newline{\bf   
    Shi-Jie Xiong$^{1}$, G. N. Katomeris$^{2}$ and \ S. N. Evangelou$^{2}$}
    \vspace{.4cm}\newline\small
   $^1$ Laboratory of Solid State Microstructures and Department of Physics,
   Nanjing University, Nanjing 210093, China \\ 
   $^2$Physics Department, University of Ioannina, Ioannina 45110, Greece \\
    \vspace{.4cm}\newline\begin{minipage}[h]{\textwidth}\baselineskip=10pt
    {\bf  Abstract.}
In the absence of magnetic field or spin-orbit coupling 
the one-parameter scaling theory predicts localization 
of all states in two-dimensional ($2D$) disordered 
systems, for any amount of disorder.  
However, a $2D$ metallic phase has been recently reported
in high mobility Si-MOS  and GaAs-based materials
without  magnetic field. We study numerically a
recently  proposed  $2D$ model which consists of
a compactly coupled pure-random plane
structure. This allows to obtain exactly
a continuum of one-dimensional ballistic extended states 
which can lead to a marginal metallic  phase 
of finite conductivity $\sigma_{0}=2e^2/h$, 
in a wide energy range  whose boundaries define 
the mobility edges of a first-order metal-insulator transition.
We present numerical diagonalization results
and the conductivity of the system 
in perpendicular magnetic field,
which verify the above analytical predictions. 
The model is also discussed in connection
to recent experiments.
    \end{minipage}\vspace{.4cm} \newline {\bf  Keywords:}
Localization, $2D$ Metal, Metal-insulator transition 
    \newline\vspace{.2cm} \normalsize

\section{Introduction }

The research activity in the study of electronic transport 
in disordered mesoscopic systems has increased dramatically
in recent years, following the rapid advances in microstructure 
technology \cite{ALW91}. Anderson localization of eigenstates 
and the associated metal-insulator transition (MIT)
is the important phenomenon in this area, which is
due to the presence of disorder  \cite{Anderson58}. 
Accordingly, a scaling theory was developed 
to deal with the Anderson MIT, 
which predicts that all states are exponentially localized
in two--dimensional ($2D$) disordered systems \cite{abr79}. 
Extended states with metallic diffusive 
regime can exist in higher than two dimensions,
e.g. in weakly disordered $3D$ systems. A dramatic 
experimental breakthrough in this
area concerns the Si-MOSFET microstructures where 
a $2D$ metallic phase was demonstrated
at zero magnetic field. This unusual finding 
was obtained by  studying the Quantum Hall Effect
to insulator transition where the extended states, 
centered in the middle of each Landau band
at high field $H$, were shown to coalesce and remain 
in a finite energy range as $H$ approaches zero \cite{Kr951}.
This behavior clearly suggests a metallic phase and
a $2D$ metal-insulator transition in 
contradiction with the scaling theory of localization
which predicts that the extended states should `float up' 
in energy as $H\to 0$, leading to complete localization.

The main experimental findings can be summarized 
in the  strong exponential drop of the temperature dependent 
resistivity at temperatures below $2K$, 
in high mobility (weakly disordered)
Si-MOS samples. This exponential drop is even more
enhanced in higher mobility samples and can be
approximated by the 
empirical fit \cite{Kr95}
\begin{equation}
\rho(T) \sim \rho_{0} + \rho_{1} exp(-T_{0}/T).
\label{rho}
\end{equation}
Moreover, the obtained resistivity as a function of the electron 
density $n$ was found to be 
roughly inversely proportional to the 
distance from the critical density $n_c$
\begin{equation}
\rho(n) \sim const. /(n-n_c),
\end{equation}
with the product of the critical exponents $z\nu \approx 1$ 
obtained from the sample dependent temperature parameter
$T_0$ \cite{Kr95}, which scales as
\begin{equation}
T_0 \sim (n-n_c)^{z\nu}.
\end{equation}
The  localization length divergence
$\xi \sim |E-E_c|^{-\nu}$
at the mobility edge $E_c$  defines the localization exponent
$\nu$, and the temperature dependence of 
the inelastic scattering length $L_{in} \sim T^{-1/z}$ the 
dynamical exponent $z$, respectively. The obtained
experimental data very roughly give $\nu \approx z \approx 1$
for the two critical exponents \cite{Kr95}. 
It should be emphasised that the 
$2D$  metallic phase was subsequently
shown for n- and p-SiGe quantum wells, 
p-GaAs, p-AlAs and n-GaAs. Most data are
rather successfully interpreted by two-parameter scaling \cite{Kim}, 
since the critical conductance  varies from sample to sample,
although  it is rather close to the universal value $e^{2}/h$ 
\cite{Kr95}.

A considerable progress has been achieved in
the development of numerical techniques 
which have been particularly successful for studying Anderson 
localization. The  MIT from the localized (insulating) to 
the extended (metallic) regime can be well
understood by finite size
studies, via transfer matrix products,
and also matrix ensemble 
diagonalization, via energy level-statistical methods. 
It is desirable to perform a similar study for the
recently proposed model of a mixed (pure-random)
coupled $2D$ system \cite{Subl}, in order to demonstrate whether
disorder in one sublattice only (the other is pure)
still gives results in
agreement with the predictions of the 
scaling theory for $2D$. In this system we shall derive a
continuous $1D$ manifold of ballistic extended states
emerging, within the majority of localized states.
The presence of this manifold of extended states will be 
also confirmed by numerical studies of the 
corresponding tight-binding Hamiltonian
in the presence of disorder $W$ in one sublattice
and interlattice coupling $t$. 
The localization properties of the system
will be studied by the participation ratio
of the corresponding eigenstates.
Our  aim is to show that such mixed structures 
can give rise to extended states, which are required
in order to obtain a metallic phase in $2D$ disordered systems. 
The obtained results 
are surprisingly close to the experimental data 
for Si-MOS, which  indicate some kind of a
first-order MIT \cite{Dob97}.

The paper is organized as follows:
Section II gives an overview of the model and the
physical parameters characterizing the proposed
mixed system. In Section III we 
derive the $1D$ extended states.
The density of states and the eigenstate behavior
of the model is studied by numerical diagonalization 
in Section IV,
whereas the properties of the conductance in perpendicular 
magnetic field in Section V. 
Finally, in the last section we give a summary of our
conclusions.

\section{The model}

Although the one-parameter scaling theory 
predicts always localization
in one or two dimensions,
isolated extended states can exist even 
in $1D$ as a consequence of
specific short-range disorder correlations \cite{Dun90,Xi96}.  
In $2D$ disordered systems delocalization is  possible 
for noninteracting electrons  only by breaking the 
time-reversal  symmetry, via a magnetic field, 
or the spin-rotation invariance, by spin-orbit 
coupling \cite{Evan95}.
In continuous $2D$ models 
extended states were previously shown 
in a model with random $\delta $-function potentials 
of infinitesimal action distance \cite{Azb45}. 
The presence of extended states in $2D$ random systems
is undoubtetly interesting in connection with
the  recently reported 
metallic phase and  the MIT in Si-MOS structures 
and GaAs-based materials, in the absence 
of magnetic fields \cite{Kr95}. 

\begin{figure}[tbh]
\epsfxsize=3in
\epsfysize=3in
\epsffile{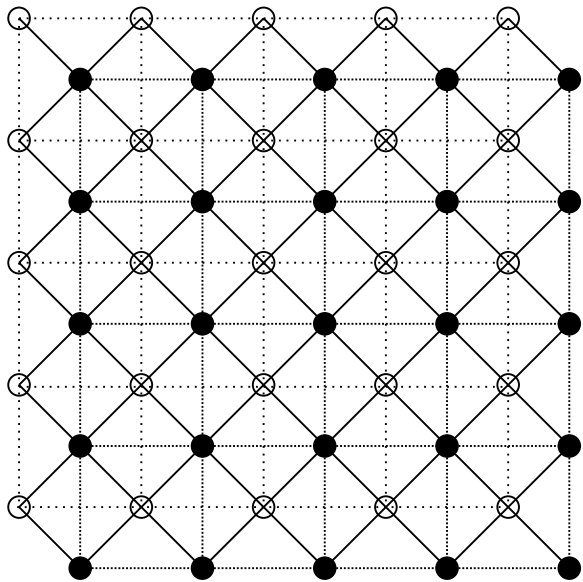}
\vspace{3mm}
\caption[fig1]{\label{fig1} \small The 
studied $2D$ pure-random coupled structure for $L=10$. 
The $L^{2}/4$ open circles denote the  pure lattice sites 
with disorder $W=0$  and the $L^{2}/4$ closed circles 
the random lattice sites with finite disorder
$W$. The two sublattices are coupled by hopping 
matrix elements $t$.}
\end{figure}

We consider the $2D$ random system shown in Fig. 1,
which is composed of two compactly 
coupled square lattices, 
one random and the other periodic.
The sites of one lattice are
located above the centers of the plaquettes 
of the other. The pure plane has zero site disorder 
and the random plane has finite site disorder estimated by 
$W$, while the coupling between the two planes is denoted
by $t$. We shall demonstrate analytically that extended  
wavefunctions can exist in the considered partially random 
structure. 
These are perfectly extended (ballistic) states with a
special momentum and form
a $1D$ continuum, which gives a finite $2e^{2}/h$ 
conductivity in a specific energy range with its
boundaries defining the mobility edges (ME) 
of a first-order MIT. 
Moreover, they can  coexist
with localized states, as expected 
from the scaling theory and 
under  long-range Coulomb interactions the system  
can reduce \cite{Subl} to the marginal Fermi liquid 
proposed in \cite{Var96}.

The corresponding tight-binding Hamiltonian  reads
\begin{eqnarray}
\label{ham}
H & = & H_{p} + H_{r} + H_{c},
\end{eqnarray}
where $H_{p}$,  $H_{r}$ and  $H_{c}$ denote the
parts corresponding to pure, random and the coupling between the
two sublattices, respectively. For  convenience, the length units 
and the origin of the coordinates are set so that 
the pure (random) sites count odd (even) numbers, e.g.
the odd indexes $(n,m)$ denote a pure lattice site 
and even indexes $(n,m)$ a random lattice
site, respectively. The Hamiltonian in the $2D$ coordinates
($n,m$) can be written as
\begin{eqnarray}
\label{ham1}
H_{p} & = & \sum_{n,m=odd}
( a^{\dagger }_{n,m}a_{n+2,m} + a^{\dagger }_{n,m}a_{n-2,m} 
+ a^{\dagger }_{n,m}a_{n,m+2} + a^{\dagger }_{n,m}a_{n,m-2}),
\end{eqnarray}
and
\begin{eqnarray}
\label{ham2}
H_{r} & =  & \sum_{n,m=even} \epsilon_{n,m} 
a^{\dagger}_{n,m}a_{n,m} \nonumber \\
& + & \sum_{n,m=even}
( a^{\dagger }_{n,m}a_{n+2,m} + a^{\dagger }_{n,m}a_{n-2,m} 
+ a^{\dagger }_{n,m}a_{n,m+2} + a^{\dagger }_{n,m}a_{n,m-2}),
\end{eqnarray}
with the coupling between them
\begin{eqnarray}
\label{ham3}
H_{c} & = &  
t\sum_{n,m}
(a^{\dagger }_{n,m}a_{n+1,m+1} + a^{\dagger }_{n,m}a_{n-1,m+1} 
+ a^{\dagger }_{n,m}a_{n+1,m-1} + a^{\dagger }_{n,m}a_{n-1,m-1})
\end{eqnarray}
where $a_{n,m}$
is the destruction operator for an electron 
at the site $(n,m)$ of the pure plane (n,m=odd) 
or the random plane (n,m=even).  The coupling
Hamiltonian $H_{c}$ connects  each pure (random) site to its four
nearest-neighbor random (pure) sites 
via the hopping parameter $t$. 
The site energies in the random plane $\epsilon_{n,m}$ 
are from a uniform distribution between $-W/2$ and $W/2$.

\section{Extended states in the $2D$ random structure}

Extended wave functions can have 
zero amplitude on the random sites
and finite amplitude on the pure sites. This requirement
leads to the equation 
\begin{eqnarray}
\label{tt}
t ( \psi(n,m)  + \psi(n+2,m)  + \psi(n,m+2)  + \psi(n+2,m+2) ) 
& = & 0,
\end{eqnarray}
for the wavefunction amplitudes $\psi$ 
on the four pure lattice sites with odd 
$(n,m)$, $(n+2,m)$, $(n,m+2)$ and $(n+2,m+2)$ which are 
adjacent to the random 
lattice site $(n+1,m+1)$ having zero amplitude. This equation
is satisfied by the pure plane wave states
$\psi_{k_x,k_y} (n,m) \sim \exp (ik_x n + ik_y m)$, which leads to
the condition
\begin{eqnarray}
\label{ttt}
(1 + \exp (i2k_x) (1 + \exp (i2k_y))
& = & 0,
\end{eqnarray}
that is
$k_x  = \pi/2$, or $ k_y  = \pi/2$.
Therefore, we can immediately write down extended states 
which are completely unaffected 
by the randomness  in the considered 
$L \times L$ structure, with periodic boundary conditions
in all directions. In order to obtain such states for 
finite $L$ with 
the adopted boundary conditions, $L$ must be even and multiple
of four so that $k_x = \pi/2$ or $k_y = \pi/2$ denote
eigenstates of the pure plane. They take the normalized form
\begin{equation}
\label{ext1}
\psi_{1,k_x}(n,m)=\frac{2}{L}\sin (\frac{m\pi }{2})\exp (ik_x n),
\end{equation}
\begin{equation}
\label{ext2}
\psi_{2,k_y}(n,m)=\frac{2}{L}\sin (\frac{n\pi }{2})\exp (ik_y m), 
\end{equation} 
where $n,m$ are the odd site coordinates
of the pure lattice and $k_{x}(k_{y}) = 2j_{x}(j_{y})\pi/L$, 
$j_{x}(j_{y}) = 1,2,...,L/2$. The states 
$\psi_{1,k_x}(n,m)$, $\psi_{2,k_y}(n,m)$ are exact 
eigenstates of the Hamiltonian $H$ which survive in the 
considered mixed pure-random coupled structure as it
can be easily verified by applying $H$ on 
$\psi_{1,k_x}(n,m)$, $\psi_{2,k_y}(n,m)$. One 
obtains the corresponding eigenenergies of these states
\begin{equation}
\label{ene}
E=2[\cos (2k_x)-1], E=2[\cos (2k_y)-1],
\end{equation}
respectively.  Since their transverse momentum 
is fixed to  $k_{y}(k_{x})=\pi/2$ while
their longitudinal momentum 
$k_{x}(k_{y})$ runs in the $1D$ Brillouin zone,
they form a $1D$ continuum 
in the energy range $[-4,0]$ \cite{Subl}.

\begin{figure}[tbh]
\epsfxsize=3in
\epsfysize=3in
\epsffile{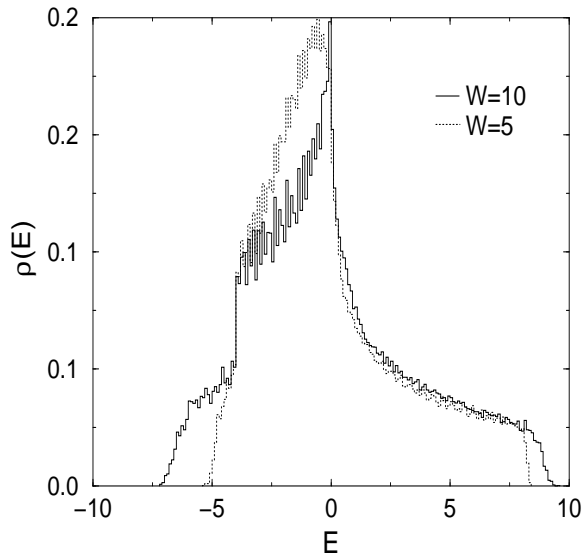}
\vspace{3mm}
\caption[fig2]{\label{fig2} \small The density of states $\rho(E)$,
in arbitrary units, for the considered $2D$ random structure
with  linear size $L=100$ with $t=1$, by taking
average over 8 random configurations. It the region $[-4,0]$ 
the continuous  manifold of pure extended states coexists with
other localized states. }
\end{figure}

The perfectly extended states found
have momentum along both the $2D$ principal axes,
$x$ (or $y$) direction, 
independently of the system size $L$. 
Moreover, they do not violate the scaling theory 
since they effectively decouple from the random lattice
although the two sublattices remain compactly coupled.
They lead to
minimum metallic conductivity $\sigma_{0}=2 e^2/h$ 
(if spin is included) and
one obtains a first-order MIT from $\sigma_{0}$ to $0$ at 
$E_{c1}=-4$ and $E_{c2}=0$, which can be regarded as 
mobility edges \cite{Subl}. The rest of the $2D$ states are
asymptotically localized, as expected from the scaling theory 
\cite{abr79}. We also observe 
that the lower ME at $E_{c1}=-4$ lies near the 
bottom of the band  so that the  
Fermi energy can be moved into the 
metallic regime even for small electronic doping,
in agreement with silicon MOSFETs 
where the metallic phase is achieved at 
very low electron densities ($\sim 10^{11} $cm$^{-2}$) \cite{Kr95}. 
It should be also noted that a minimum metallic conductivity
around $e^2/h$ has been suggested
from experimental data for holes in 
GaAs heterostructures \cite{Kr95}.

We should also notice that the obtained $\sigma=\sigma_0$ 
in the energy window $[E_{c1}, E_{c2}]$ is
scale invariant, so that
the system does not reache a {\it true} metallic phase
but a critical one instead. The corresponding 
$\beta$-function becomes zero and not 
positive in the considered $2D$
system \cite{abr79}. Moreover, the obtained extended states 
which coexist with the localized states in 
the above wide energy range, have lower dimensionality and measure, 
so that in ref. \cite{Subl} the obtained phase
was renamed ``marginal metallic phase'', adopting
the terminology ``marginal Fermi liquid'' 
proposed by Varma {\it el al} \cite{Var96}.

\section{Numerical diagonalization}

We have performed exact numerical 
diagonalization of the $L\times L$ 
($L/2 \times L/2$ pure and 
$L/2 \times L/2$ random) system adopting periodic boundary
conditions in the two directions. 
We obtain  $L/2+L/2-1 = L-1$ perfect $1D$ extended 
electronic states among the rest of 
localized states in $[-4,0]$, out of the total number
of $L^{2}/2$ states. 
In Fig. 2 we show the obtained density of states 
for two values of the disorder $W$. It is seen that the 
characteristic logarithmic spectral 
singularity of the pure lattice at $E=0$  
in two dimensions is 
recovered for large disorder $W$. This  is due 
to the fact that by increasing $W$ more and more 
states aquire very large localization lengths, 
since the electrons avoid altogether the random lattice 
and the pure lattice eventually decouples from the system
for $W \to \infty$. 

\begin{figure}[tbh]
\epsfxsize=3in
\epsfysize=3in
\epsffile{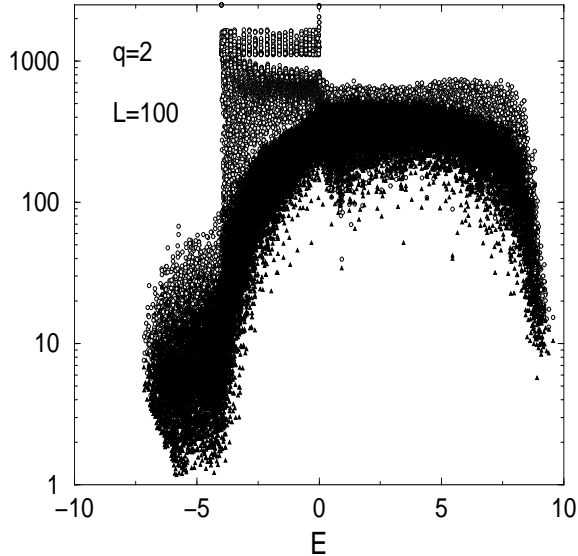}
\vspace{3mm}
\caption[fig3]{\label{fig3} \small The participation ratio $P$
(from the second $q=2$ moment of the amplitudes from Eq. (13)) 
versus energy for a fixed size $L=100$ system  with $W=10$ 
and $t=1$ by presenting data  for 8 random configurations. 
The $1D$ manifold of extended states is clearly seen
in the upper part of the figure within the energy range $[-4,0]$. 
The white(black) dots give the
participation (extend) of each state on the pure(random) 
sublattice.}
\end{figure}

We have also verified by transfer matrix studies
that the  smallest Lyapunov exponent which 
corresponds to the inverse localization length, is 
exactly zero  for any $W$ in the extended energy
window  $[-4,0]$. This calculation which confirms the presence 
of the $1D$ extended states will be presented 
elsewhere. However, from the numerical
diagonalization we can compute
the  participation ratio of each eigenstate, via
\begin{equation}
\label{par}
P= (\sum_{n,m} \Psi_{n,m}^{4})^{-1},
\end{equation}
which gives a measure of the effective number of sites occupied
by the eigenstate. $P \propto L^{2}$ for a perfectly extended
eigenstate and $P \propto const.$ for a localized state.
Fig. 3 shows the $1D$ extended states which have large $P$ values
extending over all the pure lattice. We show the
participation of the rest of states in the pure (white dots)
and the random (black dots) sublattices. The
participation is also large 
for the rest of states, despite the fact that the
disorder choice $W=10$ is the most favourable for localization
in the system. We have shown that the majority of the rest
of states have very large localization lengths and are critical
in nature, as expected in $2D$.

\section{Perpendicular magnetic field}

\begin{figure}[tbh]
\epsfxsize=3in
\epsfysize=3in
\epsffile{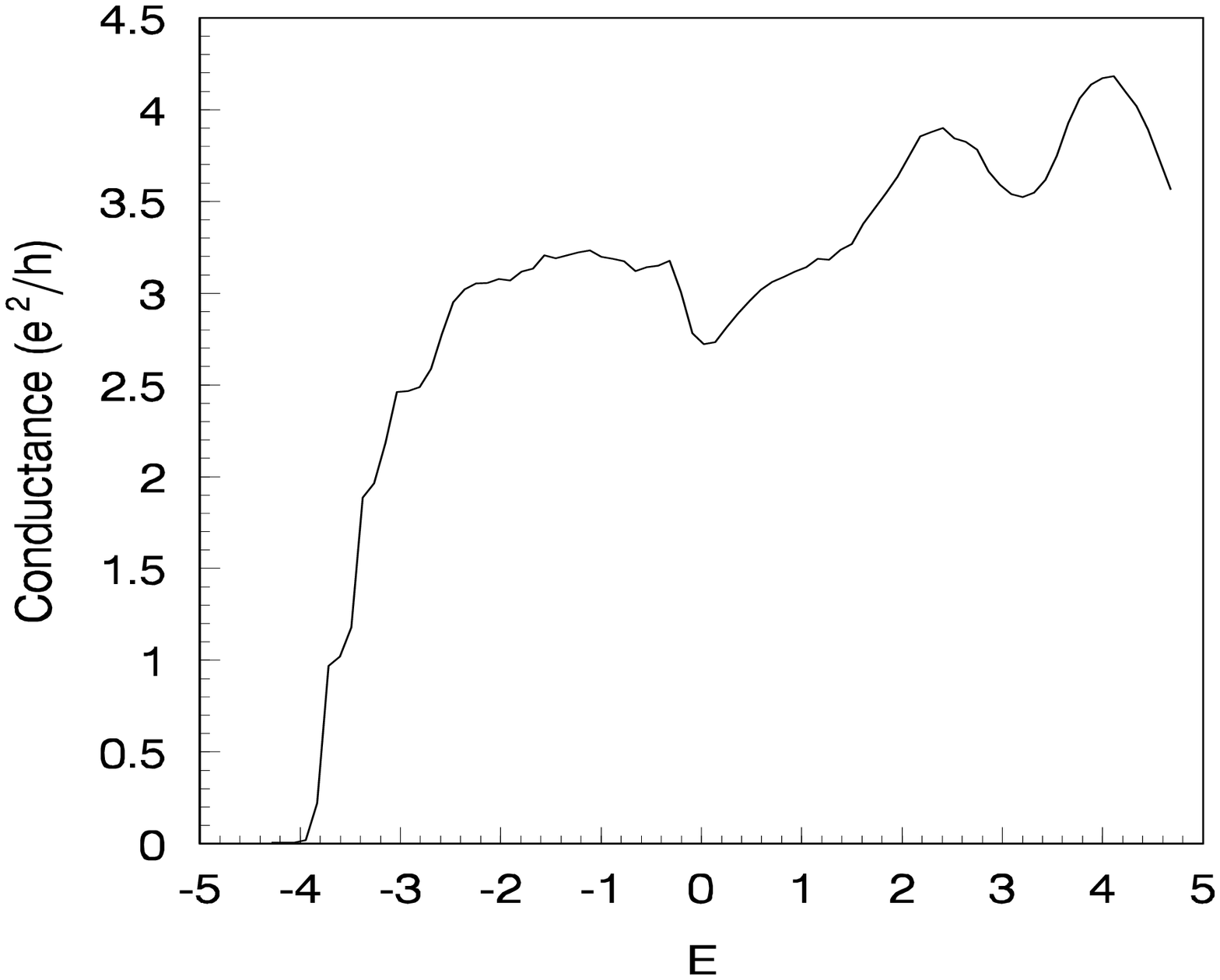}
\vspace{3mm}
\caption[fig4]{\label{fig4} \small The 
``metallic" region energy dependence
of the conductance in  a perpendicular 
magnetic field of strength of $0.1\phi_0$ 
per square plaquette, where $\phi_0$ is the flux quantum and the
fixed system size $L=24$. The 
interlattice hopping is $t=1.2$ and the disorder in the 
random sublattice is $W=8$.}
\end{figure}
 
\begin{figure}[tbh]
\epsfxsize=3in
\epsfysize=3in
\epsffile{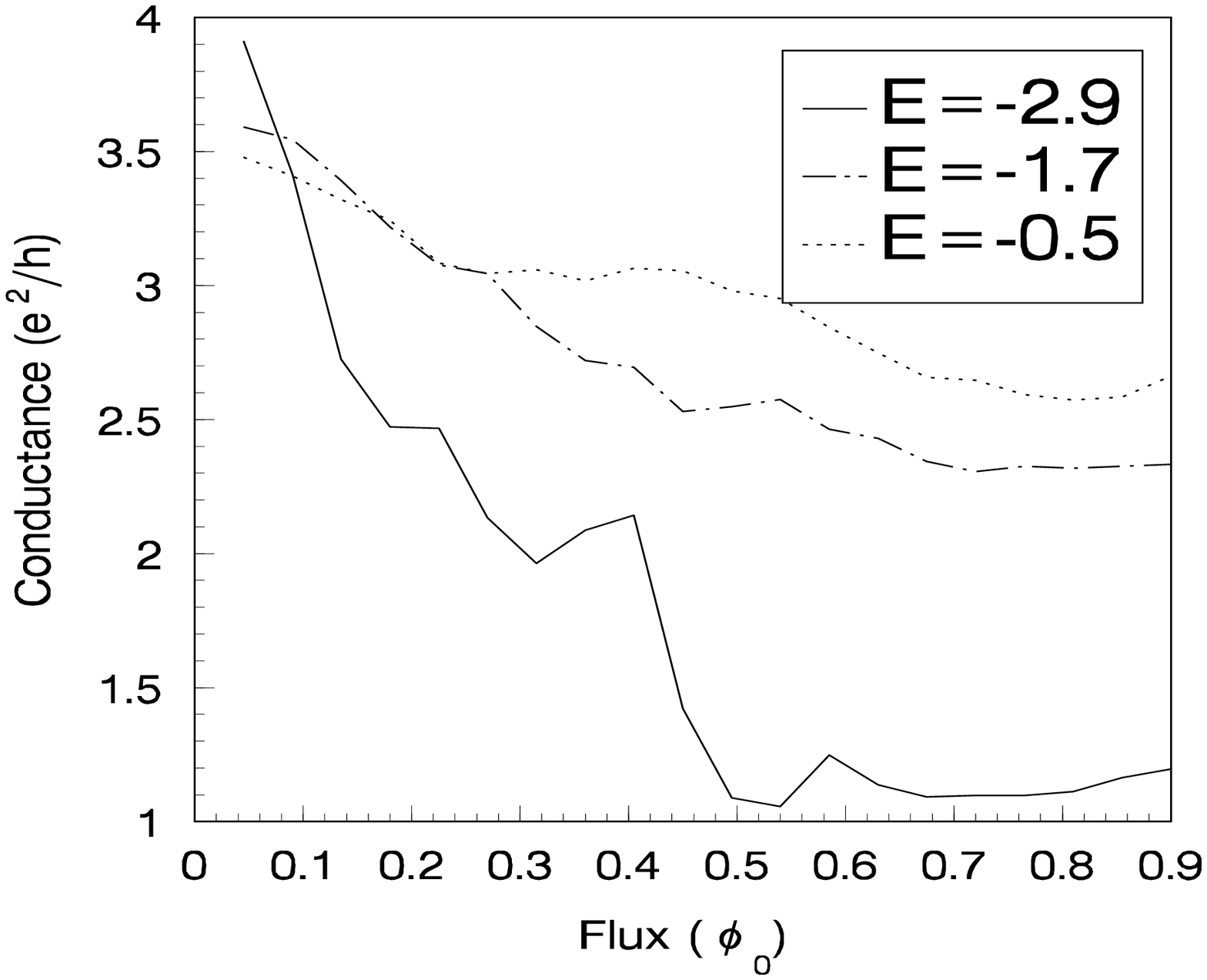}
\vspace{3mm}
\caption[fig5]{\label{fig5} \small The magnetic field 
dependence of the conductance for various 
energies in the ``metallic" regime with the
rest of the parameters the same as 
in Fig.4. We observe that the conductance drops when
the magnetic field is introduced, in agreement with
the experimental findings.}
\end{figure}

We have also computed the change of the conductance in 
the presence of a perpendicular magnetic field which 
is known to destroy the extended states \cite{Kr95}. The 
conductance $\sigma$ equals the conductivity for 
a square sample in $2D$ and can be obtained from the multichannel 
Landauer-B\"{u}ttiker formula for a $2D$ system of
size $L$ via
\begin{equation}
\sigma (L) = (e^2/h)Tr [T^{+}(L)T(L)],
\label{lb}
\end{equation}
where $T(L) $ is the $L$-channel transmission matrix for electronic
propagation. We adopt the Coulomb gauge for the  
perpendicular magnetic field  and show the obtained
results in Figs. 4 and 5. They demonstrate the reduction of the
conductance in the presence of a perpendicular magnetic field.

\section{Conclusions}

There have been a  few theoretical attempts 
in order to understand the
experimental puzzle of the $2D$ metallic phase,
such as  scaling in the presence of interactions \cite{Fil83},
via hole traps \cite{Mas},
two-phase models \cite{Xie}, 
and superconductivity \cite{Phil}, etc.
The basic mechanism was attributed to Coulomb interactions 
which leads to violation of the scaling theory 
and the presence of a metallic phase in $2D$. 
However, these approaches
suffer from serious experimental problems (e.g. 
the transition to Wigner solid, no traps or two 
phases exist for GaAs, no superconducting $T_c$, etc.)
In this paper we present
a simple transparent alternative of the $2D$ metallic phase 
via a first order MIT 
from the localized $2D$ regime to an extended
ballistic regime, which coexists with the localized phase.
For this purpose we discuss an explicitly solved model 
of a pure-random mixed structure which 
can describe the Si-MOSFET under certain assumptions.
We determine exactly the band of $1D$ ballistic extended states 
which propagate freely in the $2D$ 
mixed pure-random system, without  
magnetic field or spin-orbit coupling. 
These extended states are located at the edges of
the first Brillouin zone ($k_x=\pi/2$,  any $k_y$ or
$k_y=\pi/2$, any $k_x$) and lie in an energy regime 
whose boundaries define the first-order MIT. We 
perform numerical computation for the complete
eigensolutions of the corresponding random Hamiltonian matrix
ensemble and also present the conductance as a function of
a perpendicular magnetic field.

The extended electrons found 
form a pseudo-Fermi sea embedded in the localized electrons and
the system can be regarded as a $2D$ non-Fermi liquid because of
its measure, the $1D$-like character of the spectrum, as well as 
the presence of a background from the localized electrons. 
The pseudo-Fermi surface has only four 
points corresponding to two Luttinger liquids crossing with each 
other \cite{Subl}. Our results
are important for understanding
many experimental findings related to the $2D$ MIT
in Si MOSFETs and GaAs-AlGaAs heterostructures.
In these systems the electrons or holes are confined within 
several atomic planes and due to  lattice mismatch, 
or atomic diffusion, it could be expected that the 
planes adjacent to the interface are random, while some 
planes relatively far  from the interface are 
less random and, as we have shown,
could support ballistic channels. 
Although for these materials more realistic models are 
certainly needed
our results may still shed some light on 
the important phenomenon of the $2D$ metallic phase.  
%
    \vspace{0.6cm}\newline{\small 
This work was supported in part  by TMR and also a 
$\Pi$ENE$\Delta$ Research Grant of the Greek Secretariat of Science and 
Technology. 

    }
    \end{document}